\documentclass[10pt,preprint]{aastex}

\shorttitle{2-D Orbits in Galactic Disks}
\shortauthors{H. Nakanishi}
\begin{document}
\title{A Method For Determining Two-Dimensional Orbits In Galactic Disks}
\author{Hiroyuki Nakanishi}
\affil{Institute of Astronomy, University of Tokyo,
2-21-1 Osawa, Mitaka, Tokyo 181-0015}
\begin{abstract}
We propose a method for determining the two-dimensional orbits of interstellar matter (ISM) in galactic disks using observed line-of-sight velocity fields. Although four components, consisting of two of position and two of velocity, are necessary to determine two-dimensional orbits, the velocity perpendicular to the line of sight can hardly be determined from observations. We show that the ISM orbits can be approximately determined by assuming that (1) the angular momentum is conserved along each orbit, (2) all orbits are closed in a single rotating frame, and (3) the system is symmetric through the origin. Since the angular momentum can be estimated on the major axis of a disk, a velocity perpendicular to the line of sight can be determined according to the conservation of angular momentum, and thus two components of velocity are obtained. The ISM orbits can be calculated by solving differential equations given by an obtained two-dimensional velocity field when a certain pattern speed is given. The typical deviation from the correct orbit is less than the observational resolution according to our test calculations for modeled velocity fields. We applied this method to the observed data of NGC 4569 and obtained possible orbits for this galaxy. Furthermore, we calculated the mass distribution using our result.
\end{abstract}

\keywords{Galaxies: Kinematics and Dynamics, Galaxies: ISM, ISM: Kinematics and Dynamics, Methods: Data Analysis}

\section{Introduction}
Although orbits of stars or interstellar matter (ISM) on galactic planes are fundamental in studying galactic dynamics, accurate shapes of these in real galaxies are difficult to discern because two-dimensional motion can hardly be observed. Observations of line emission give only the line-of-sight velocities of the ISM in spiral galaxies. Hence, we often discuss galactic dynamics assuming that the ISM motion is circular, since the velocity fields of galaxies often show so-called spider diagrams, which can be reproduced by circular rotation around their galactic centers. Obtained rotation curves provide mass distributions in spiral galaxies (Sofue et al. 1999, 2003a). Barred galaxies, however, show strong noncircular motions, and approximations of circular orbits fail. For these galaxies, the computer simulation of ISM dynamics has played an important role in studying galactic dynamics, and many models have reproduced observations fairly well (Huntley 1978; Wada et al. 1998). However, the link between simulations and observations is generally difficult, since there are many unidentified parameters, such as rotation curve, bar strength, pattern speed, viscosity, etc. And although the shape of a galactic bar is often expressed in the form $\Phi_b(R) \cos{(m\phi)}$, where $R$, $m$, and $\phi$ denote galactic radius, integer, and the angle around the galactic center, respectively, it has not been determined by observations.

     Therefore, it is desirable to determine two-dimensional ISM motions on the galactic plane directly from observations in order to model real galaxies. While Kuno \& Nakai (1997) inferred an ISM orbit by tracing the velocity vectors at the spiral phase using the CO data of M51 and found that the orbit deviates greatly from a purely circular motion, there are few galaxies for which two-dimensional orbits have been derived from observational data.

     In order to trace two-dimensional motion in a galactic disk we need four values, two of which are of position and two of which are of velocity. Both position components are directly given by observation. A velocity component perpendicular to the line of sight cannot be determined, although the line-of-sight component is given by observations.

     In this paper, we propose a method for determining the two-dimensional orbits of ISM on galactic planes directly from observations on the basis of simple assumptions. We applied this method to model velocity fields to confirm that it works well. We then applied this method to a real galaxy, NGC 4569, for which $^{12}$CO($J=1-0$) data has been obtained with Nobeyama Millimeter Array (NMA) observations. 

\section{Method}
When the potential of the galactic disk is considered to be dominated by an axisymmetric component and bar or spiral components are small enough, the force acting on the ISM is directed approximately centripetally and the angular momentum is nearly constant. In this section, we describe a method for determining an ISM orbit on a galactic plane based on the assumptions that (1) the angular momentum is conserved along an orbit, (2) all orbits are closed in a single rotating frame, and (3) the system is symmetric through the origin.

     We choose a Cartesian coordinate whose ($x, y$)-plane coincides with a galactic plane. In viewing the galaxy face-on, we let the $x$- and $y$-axes coincide with the major and minor axes of the disk, respectively. The galactic center is located at ($x, y$) = (0,0). The velocity components parallel to the $x$- and $y$-axes are denoted by $u$ and $v$, respectively. When the galactic disk is inclined by $i$, the projected coordinate is defined by $x_{\rm proj}$, $y_{\rm proj}$, $u_{\rm proj}$, $v_{\rm proj}$, where $x_{\rm proj} = x$, $y_{\rm proj} = y \cos{i}$, $u_{\rm proj} =u$, $v_{\rm proj}=v\sin{i}$. The observable line-of-sight velocity corresponds to $v_{\rm proj}$. While three components $(x,y,v)$ are known from observation with the inclination $i$, one component, $u$, is unknown.

     Let a galaxy rotate counterclockwise and the lower side of the disk ($y < 0$) be near us. First, we choose two points on the major axis so that they are symmetric with respect to the galactic center; these points are defined as $(x_{0}^1, y_{0}^1) = (a, 0)$ and $(x_{0}^2, y_{0}^2) = (-a, 0)$. The superscript denotes a quadrant of position; 1, 2, 3, and 4 mean the first, second, third, and fourth quadrants, respectively. The subscript represents position; 0 denotes an initial point and 1 denotes a neighboring point. Here we mainly think of two trajectories in the first and the second quadrants, that is, $y > 0$. We omit a superscript when we do not specify the quadrant.

     The velocities $v_{0}^1$ and $v_{0}^2$ at these points are determined from observation. We set $v_{0}^1=-v_{0}^2=v_0$ by averaging the observational data at two points in order to reduce observational errors. We suppose that one orbit crosses these two points and that the angular momentum, $L$, is constant along this orbit. Since the angular momentum in the Cartesian coordinate is
\begin{equation}
L=x v - y u
\end{equation}
and $y = 0$ along the major axis, the angular momentum along this orbit is determined uniquely as $L = a v_0$. Since $v$ at each point $(x, y)$ is given by observation, $u$ can be estimated to be
\begin{equation}
u = (x v - L)/y
\end{equation}
at each point.

     Second, we set the velocity $u_{0}$ to be $v_{0}/\tan{\theta}$ at $(x_{0}, y_{0})$ using a free parameter, $\theta$. Then a neighboring point $(x_{1}, y_{1})$ is determined in the vector form approximately by
\begin{equation}
{\bf x}_{1} = {\bf x}_{0} + ({\bf v}_{0} - {\bf v}_{\rm pat}) dt, 
\end{equation}
where ${\bf x}$ and ${\bf v}$ denote position and velocity vectors, respectively. The vector ${\bf v}_{\rm pat}$ is the pattern speed, whose absolute value is proportional to the radius and whose direction is always tangential to the position vector. The constant $dt$ is the time interval. This equation is a formula of the Euler method for solving a differential equation numerically. The time interval $dt$ is chosen so that each subsequent point is separated from the initial point by half the spatial resolution of the observation, based on the Nyquist sampling. We allow $dt$ to take a negative value; positive and negative $dt$ correspond to trajectories traveling counterclockwise and clockwise from the initial points. Two trajectories, therefore, start from one initial point.

     The vector component $v_{1}$ at the next position $(x_{1}, y_{1})$ can be determined by observation, and the other component, $u_{1}$, may be estimated by the conservation of angular momentum. We calculate an orbit in the first quadrant simultaneously with the third quadrant, which is the same as in the second and fourth quadrants. Then, to reduce the observational noises, we average velocity components $v$ at symmetric points relative to $(0,0)$ every calculation step; for example, ${\bf v}_{1}^{1} = ({\bf v}_{1}^{1} - {\bf v}_{1}^{3})/2$. If data is lacking at the next point, we define the value at this point to be the same as that of the previous; ${\bf v}_{k} = {\bf v}_{k-1}$, where $k$ is any integer. We repeat this process until an orbit reaches $x=0$. For the next stage, we vary a parameter $\theta$ by $\Delta \theta$, which was taken to be $1\arcdeg$ in this paper, so that orbits from $(\pm a, 0)$ meet one another at $x=0$. We avoid a solution whereby a radius of the orbit becomes extremely large, $|{\bf x}_{k}| > r_{\rm upper}$. In this paper we adopted $r_{\rm upper} = 5 x_{0}$.

     Figure 1 presents a flow chart and a scheme for this method. Although we explain our method by using the Euler method for simplicity, we actually use a fourth-order Runge-Kutta method to calculate orbits more accurately.

\vskip 3mm
\centerline{--- Figure 1   ---}
\vskip 3mm

\section{Application To Observed Data}

\subsection{Data}
In this paper we applied this method to NGC 4569 $^{12}$CO($J=1-0$) data, which was taken from the high-resolution Virgo CO survey (Sofue et al. 2003b). The adopted parameters and observational parameters are listed in Table 1. The inclination and position angles were determined by applying an elliptical fitting to a K-band image (Jarrett et al. 2003) using the task ellipse in the IRAF STSDAS software. The position angle steeply decreases toward the center within $R<3\farcs5$, which reminds us of a central bar. The position angle largely increases outward beyond $R>131\arcsec$, which reminds us of a warping of the stellar disk. In order to avoid these effects, we estimated inclination and position angles by averaging them between $R=3\farcs5$ and $131\arcsec$. Consequently, we obtained the position angle $18\arcdeg$ and the axial ratio $q=0.52$, which yields an inclination of $i=63\arcdeg$;, according to the equation of Tully (1988),
\begin{equation}
i=3\arcdeg + {\rm cos}^{-1} \sqrt[]{(q^2-0.2^2)/(1-0.2^2)}. 
\end{equation}
NGC 4569 shows an S-shaped velocity field, which indicates the clear existence of noncircular motion because the isovelocity contour of the systemic velocity is not parallel to the minor axis at all. The Digitized Sky Survey (DSS) $B$-band and CO images of this galaxy are shown in Figures 2a and 2b. The contours in Figure 2b represent the velocity field.

According to Tremaine \& Weinberg (1984), the pattern speed can be estimated by using the intensity-weighted line-of-sight velocity and the position of the major axis based on the equation of continuity. We applied this method to this galaxy using the $K$-band image and the CO velocity field. The intensity at the center is quite large, which reminds us of nuclear activity and does not seem to satisfy the equation of continuity. Therefore, we estimated the pattern speed in the range $2\arcsec \leq R \leq 9\arcsec$ and obtained a pattern speed of 74 km s$^{-1}$ kpc$^{-1}$. Note that this value is tentative, since gaseous velocity does not always coincide with the stellar velocity.

\vskip 3mm
\centerline{--- Table 1   ---}
\vskip 3mm

\vskip 3mm
\centerline{--- Figure 2   ---}
\vskip 3mm

\subsection{Results}
We applied our method to this galaxy to determine orbits on the galactic planes. For NGC 4569, we fixed the step of calculation at $1\farcs55 $, which is half of the beam size. We calculated less than 12 orbits that cross the points $(x,y)=(\pm 0\farcs775 n, 0)$, where $n$ is an integer, on the major axis. We show the resultant orbits viewed face-on in Figure 2c. We imposed these orbits on the CO image, which is shown in Figure 2d. In NGC 4569, orbits tends to be strongly elongated in the direction of the major axis. The axes of the orbits are slightly inclined in the counterclockwise direction from the major axis. This large deviation from circular orbits causes the strong disturbance in the velocity field.

     Since the estimated pattern speed is tentative, we calculated orbits taking other pattern speeds (34, 54, and 94 km s$^{-1}$ kpc$^{-1}$) to find how the resultant orbits depend on the pattern speed. Results are shown in Figure 3. Fewer orbits can be obtained when a larger pattern speed is adopted. In any case, resultant orbits are strongly elongated in the direction of the major axis and are inclined in the clockwise direction from the major axis.

\vskip 3mm
\centerline{--- Figure 3   ---}
\vskip 3mm

Since characterizing the presence of noncircular motions is sensitive to the values of inclination and position angle adopted (Schoenmakers et al. 1997), we calculated orbits taking other inclinations and position angles that differ by $3\arcdeg$ from $i=63\arcdeg$ and $P.A. = 18\arcdeg$. Results are shown in Figure 4. Only a few orbits can be obtained for some adopted inclinations and position angles. In these cases also, resultant orbits are elongated and are inclined in the counterclockwise direction. 

\vskip 3mm
\centerline{--- Figure 4   ---}
\vskip 3mm

While some orbits seem to cross others slightly, the gas orbits should not cross each other, otherwise the gasses would collide, shock, and lose angular momentum. Since the amplitude of the orbits' intersections is less than beam size, these intersections are thought to be due to a calculation error.

\section{Error Estimation}
Below we have applied our method to model velocity fields in order to confirm that it works properly and to estimate errors. Consequently, we conclude that our calculation works well within a beam size and that the calculation gives the best solution when we adopt a half beam size or a beam size as a step size.

\subsection{Model 1: Angular Momentum Is Conservative And Orbits Are Aligned}
First, we made a simple model in which each orbit is an ellipse whose axial ratio is 2 and the position angle of each orbit is the same as the others. The velocity at the major axis is fixed at 200 km s$^{-1}$, and the angular momentum is thought to be constant along each orbit. The inclination of the disk is fixed at $60\arcdeg$. The velocity field is given in an infinite resolution. When step sizes of $2\arcsec.88$ and $0\arcsec.288$ are adopted, typical deviations from the correct orbits are less than 1/20 of a step size. We conclude that the calculation works well for this ideal case.

\subsection{Model 2: Angular Momentum Is Conservative And Orbits Are Misaligned}
Next, for a more complicated case, we made a model in which each orbit is an ellipse whose axial ratio is 2 and the position angle of each orbit increases with radius; the semimajor radius is $0.\arcsec 288 n$ ($n$ being an integer), and the position angle increases by $0.\arcdeg 5$ when the semimajor radius increases by $0.\arcsec 288$. Angular momentum is constant along each orbit. The inclination of the disk is fixed at $60\arcdeg$. Velocity fields are given by smoothing with a Gaussian whose full width at half-maximum (FWHM) is $2.\arcsec 88$ on the basis of a typical beam size of the NMA Virgo CO survey. We calculated orbits with various step sizes. An averaged deviation between $2.''88$ and $10''$ gets the smallest value, $0.\arcsec 9$, when we adopt a beam size as a step size.

\subsection{Model 3: Angular Momentum Is Not Conservative And Orbits Are Misaligned}
It is not precise to assume that angular momentum is constant along an orbit in a field in which the gravitational force is not always centripetal. To see the effect of the variation of angular momentum along an orbit, we applied our method to a model of orbits in a weak bar potential. We took a model from Wada (1994). We chose a bar strength $\epsilon = 0.05 $ and a damping rate $\Lambda = \lambda/\kappa_0 = 0.1$. The angular momentum varies by $\sim 10\%$. We assume that the rotational velocity reaches a maximum of 203 km s$^{-1}$ at a radius of 1.6 kpc. The pattern speed, $\Omega_{\rm pat}$, is fixed at 70 km s$^{-1}$ kpc$^{-1}$. The distance is taken to be 16.1 Mpc, which is the distance to the Virgo Cluster. The averaged deviation is minimized at $0.\arcsec5$ when we adopt a step size equal to a half of the beam size.

     If the angular momentum varies by $a\%$, a velocity vector has an error of $a\%$ at maximum. When we fixed the step size at half a beam size, the position error of the orbit is $a\%$ of half a beam size at maximum for each step. An error due to the assumption of a constant angular momentum would not be negligible when the step number is large.

\subsection{Step Size And Error}
We found that deviation is the smallest when the step size is fixed at a beam size or a half beam size. This is interpreted as follows: The deviation grows as the step number increases. On the other hand, the smaller the step size is, the more accurately the Runge-Kutta integrator works. Since the observed data are smeared with the beam size, we cannot get information for a scale smaller than the beam size. When the step size is smaller than the beam size, the calculation error only increases. Therefore, the step number and step size should be as small as possible and the step size should not be much smaller than the beam size in order to reduce the error. Consequently, the deviation is the smallest when we adopt a beam size or half a beam size as the step size.

\subsection{Dependency On The Pattern Speed}
In order to investigate the dependency on the adopted pattern speed, we applied our method to model 3, assuming incorrect pattern speeds of 30, 50, and 90 km s$^{-1}$ kpc$^{-1}$. The averaged deviations are  $1.\arcsec5$, $0.\arcsec7$, and $0.\arcsec 9$, respectively. The deviation increases for an incorrect pattern speed. The averaged deviation is, however, smaller than the beam size for this model.

\section{Mass Distribution} 
One of the most useful applications of our method is to determine the mass distribution in a spiral galaxy even if the rotation curve cannot be obtained because of strong noncircular motion. Following Binney \& Tremaine (1987), we introduce a frame rotating at  ${\bf \Omega}_{\rm pat} = (0, 0, \Omega_{\rm pat})$ and effective potential 
\begin{equation}
\Phi_{\rm eff} = \Phi - {1\over 2} \Omega_{\rm pat}^2 R^2, 
\end{equation}
where $\Phi$ is a potential in a rest frame and $R$ is a radius. In this frame, the equation of motion becomes 
\begin{equation}
\ddot{\bf r} = \nabla \Phi_{\rm eff} - 2({\bf \Omega}_{\rm pat} \times \dot{\bf r}). 
\end{equation}
Here the acceleration $\ddot{\bf r}$ is known from the resultant orbits and two-dimensional velocity vectors. A mass density can be calculated according to the Poisson equation
\begin{equation}
\nabla^2 \Phi ({\bf r})= 4\pi G \rho ({\bf r}). 
\end{equation}

We found that the gravitational vector is oriented to the galactic center within $10\arcdeg$ by calculating it from the obtained orbits. Therefore, the potential is nearly spherical and is expressed as
\begin{equation}
\partial^2 \Phi({\bf r}) / \partial x^2 \sim \partial^2 \Phi({\bf r}) / \partial y^2 \sim \partial^2 \Phi({\bf r}) / \partial z^2
\end{equation}
 with a function of only radius $r$. Therefore, we assumed 
\begin{equation}  
\partial^2 \Phi({\bf r}) / \partial z^2 \sim (\partial^2 \Phi({\bf r}) / \partial x^2 + \partial^2 \Phi({\bf r}) / \partial y^2)/2
\end{equation}
in order to calculate the mass distribution in NGC 4569 using our result. We calculated the mass distribution with pattern speeds of 34, 54, 74, and 94 km s$^{-1}$ kpc$^{-1}$. 

     Figure 5 shows the mass distribution on the galactic plane of NGC 4569. In this paper, we calculated mass densities inside the largest obtained orbits. NGC 4569 shows an elongated distribution which is inclined in the counterclockwise direction from the major axis. The mass distribution near the borderlines is disturbed because there are not many data points to calculate.

A typical error arising from a 10.4 km s$^{-1}$ error in the line-of-sight velocity is 5.7 $M_\odot$ pc$^{-3}$ when we suppose (1) a velocity vector to be parallel with $y$-axis; (2) an error in a gradient of the potential to be $\delta |\nabla \Phi| \sim  2\delta v / \delta t + 2{\Omega}_{\rm pat} \delta v \sim 2 v \delta v / \delta y + 2{\Omega}_{\rm pat} \delta v$; and (3) an error in mass density to be $4\pi G \delta \rho \sim  2\delta |\nabla \Phi| / \delta y $. Here, we adopt $\delta v = 10.4$ km s$^{-1}$, $\Omega_{\rm pat} = 74$ km s$^{-1}$ kpc$^{-1}$, $v = 100$ km s$^{-1}$, and $\delta y =  1.55\times 78$ pc.

\vskip 3mm
\centerline{--- Figure 5   ---}
\vskip 3mm

\section{Remarks In Applying Our Method}

We presented the ISM orbits as being approximately determined in assuming that (1) the angular momentum is conserved along each orbit, (2) all orbits are closed in a single rotating frame, and (3) the system is symmetric through the origin.
Our method is useful for modeling a real galaxy because it is calculated directly from observational data. However, we must note the following points: First, our method is plausible for a potential that is dominated by an axisymmetric component and in which the loss of angular momentum due to the shock of the gas is negligible. Otherwise, the angular momentum is not conserved and our method fails. Second, this method cannot be applied to a patchy velocity field. Third, this method cannot work at the corotation radius where the rotational velocity is equal to the pattern speed. Fourth, resultant orbits can deviate by observational resolution.

\acknowledgments
     We would like to thank Yoshiaki Sofue and Akihiko Ibukiyama for invaluable comments and advice on writing the manuscript. We are grateful to Sachiko Onodera and Keiichi Wada for useful comments. We also thank the anonymous referee for useful comments and suggestions. H. N. is financially supported by a Research Fellowship from the Japan Society for the Promotion of Science for Young Scientists.


\begin{table}
 \caption{Adopted Parameters and Observational Parameter}
  \begin{tabular}{cc}
   \hline\hline
Name                                  & NGC 4569 \\
R.A.(J2000)&$12^{\rm h} 36^{\rm m} 49.80^{\rm s}$$^{*1}$\\
Dec.(J2000)&$+13\arcdeg 09\arcmin 46.3\arcsec$$^{*1}$\\
Morphology type & SAB(rs)ab$^{*2}$\\
inclination & {$63\arcdeg$$^{*3}$}\\
position angle & {$18\arcdeg$$^{*3}$}\\
systemic velocity [km s$^{-1}$]& -235 $\pm$ 4$^{*2}$\\
beam size & $4.\arcsec 5 \times 3.\arcsec 1$$^{*4}$\\
r.m.s. noise [mJy] & 25$^{*4}$ \\
velocity resolution [km s$^{-1}$]& 10.4$^{*4}$ \\
   \hline
   \hline
  \end{tabular}\\
$^{*1}$ The 2MASS Large Galaxy Atlas \citep{jar03}\\
$^{*2}$ RC3 \citep{dev92} \\
$^{*3}$this work (elliptical fitting using the task {\it ellipse} in the {\it IRAF STSDAS} software for 2MASS K-band image)\\
$^{*4}$\citet{sof03a}\\
\end{table}


\begin{figure}
\plotone{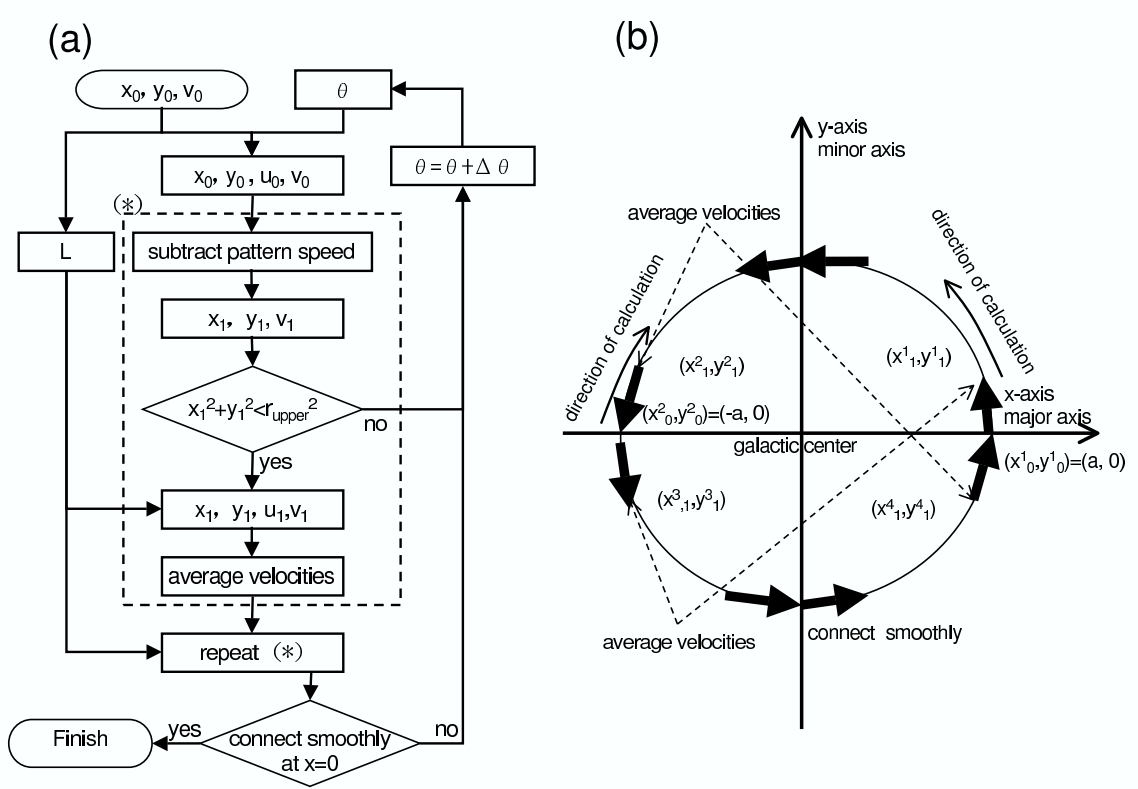}
\caption{ A flow chart (left) and a scheme (right) for our method.\label{fig1}}\end{figure}

\begin{figure}
\plotone{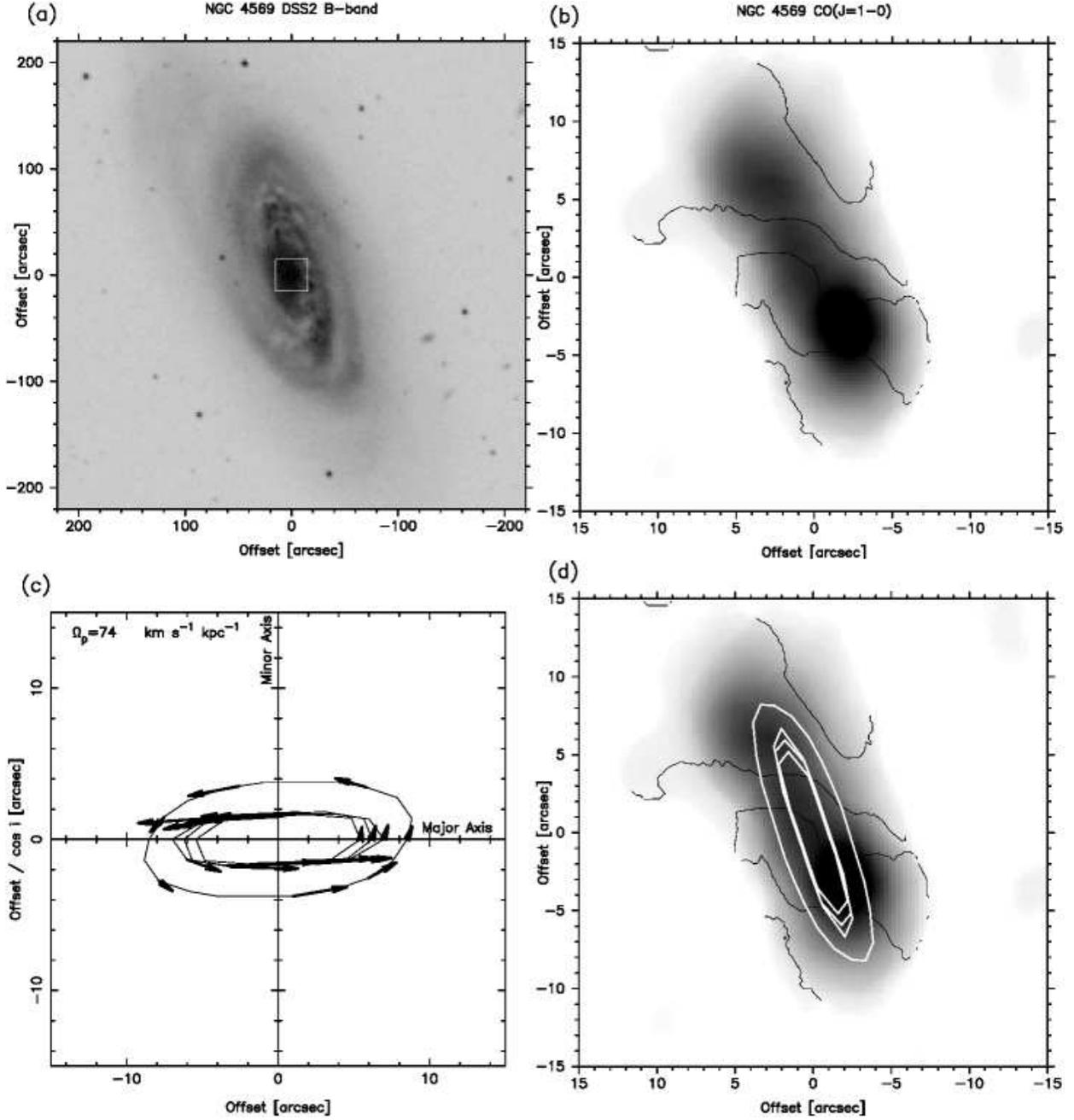}
\caption{(a) DSS B-band image of NGC 4569.  A square indicates the area of (b). (b) A $^{12}$CO (J=1-0) map with isovelocity contours; -335, -285, -235, -185, and -135km s$^{-1}$. (c) Resultant orbits and two-dimensional velocity vectors viewed face-on. (d) Resultant orbits superposed onto the CO image of (b).}
\end{figure}

\begin{figure}
\plotone{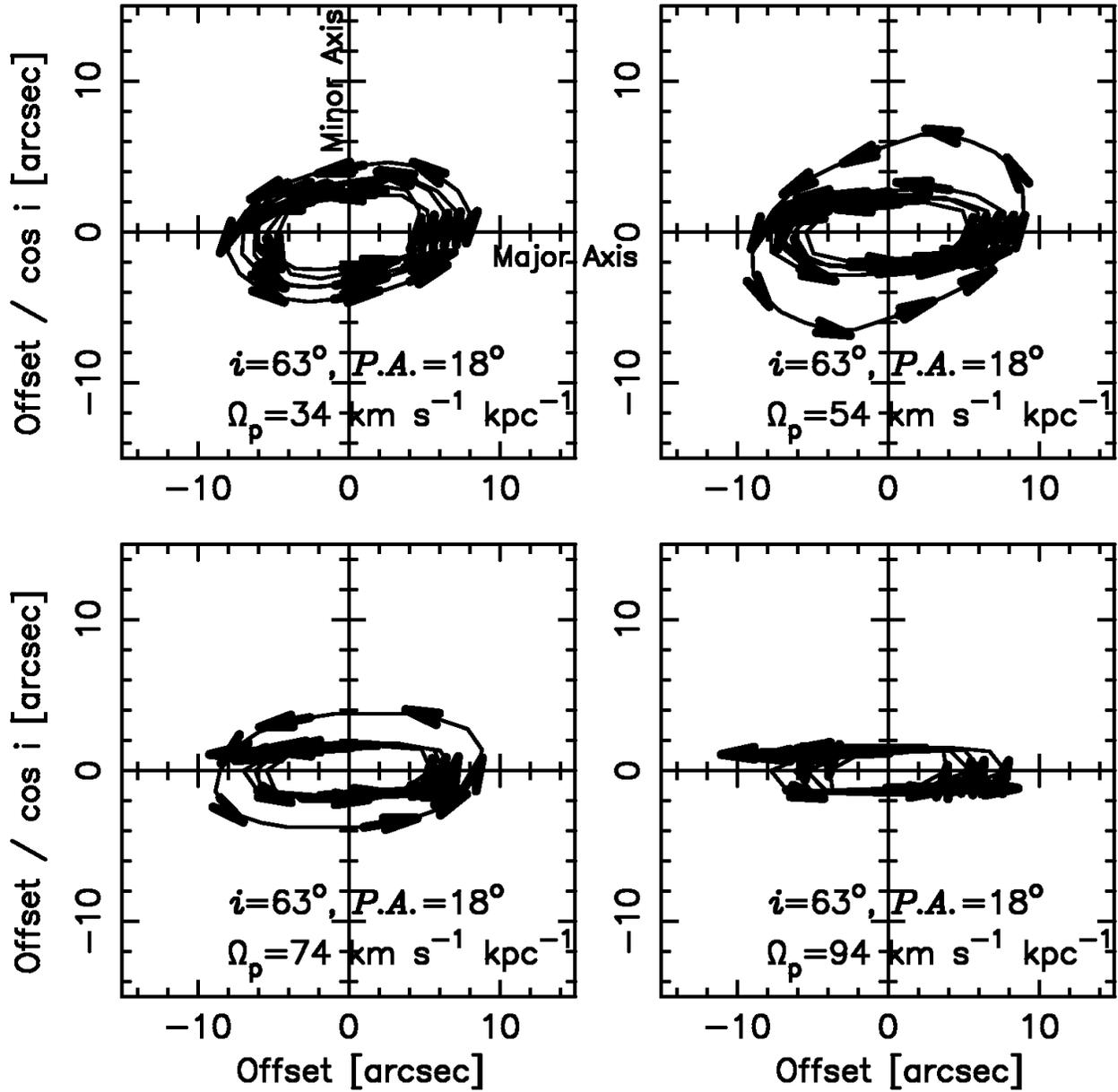}
\caption{Resultant orbits and two-dimensional velocity vectors assuming that the pattern speed is 34, 54, 74, and 94 km s$^{-1}$ kpc$^{-1}$. }
\end{figure}

\begin{figure}
\plotone{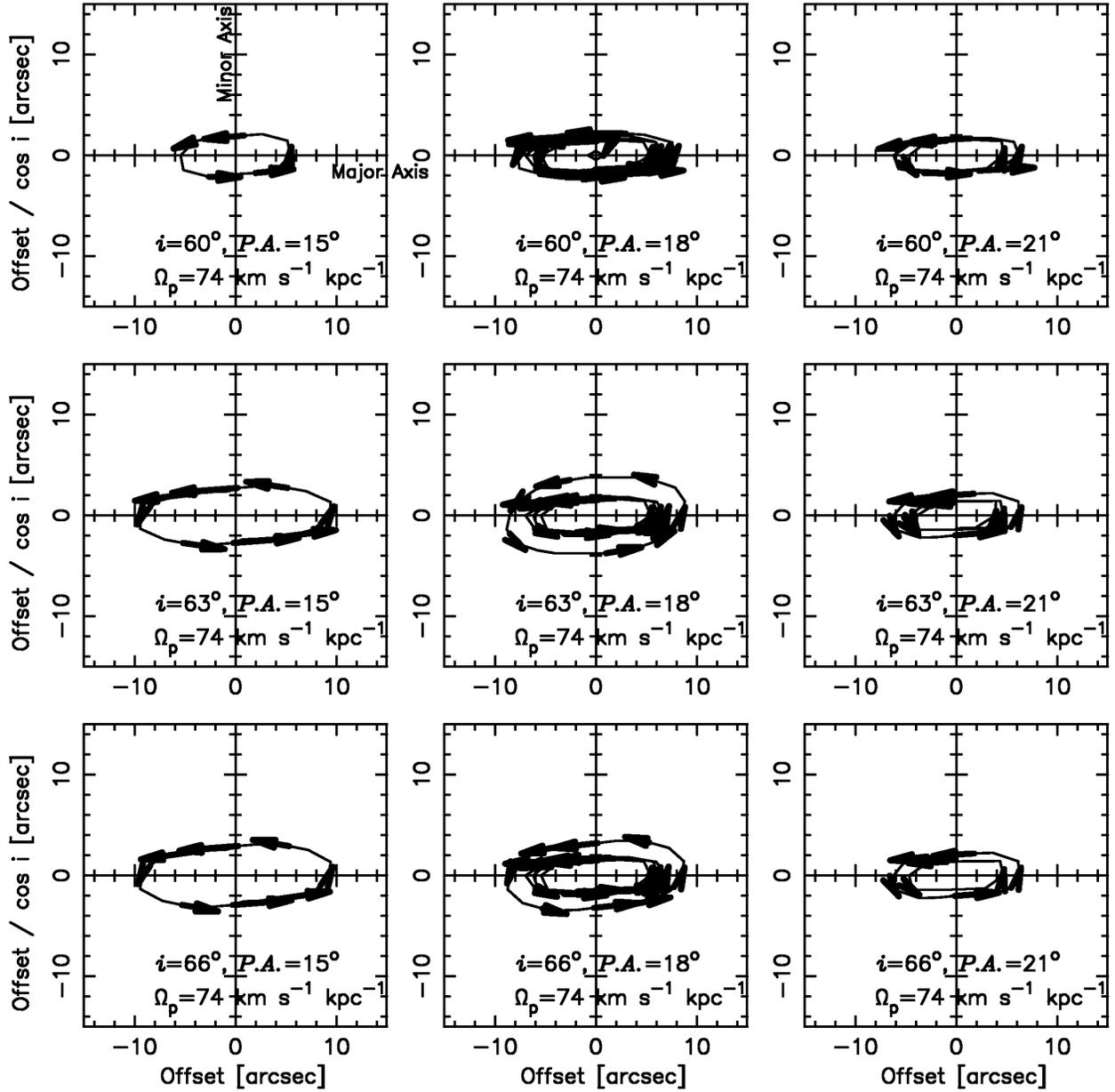}
\caption{Resultant orbits and two-dimensional velocity vectors assuming that the inclination is $60\arcdeg$, $63\arcdeg$, and $66\arcdeg$, and that the position angle is $15\arcdeg$, $18\arcdeg$, and $21\arcdeg$. }
\end{figure}

\begin{figure}
\plotone{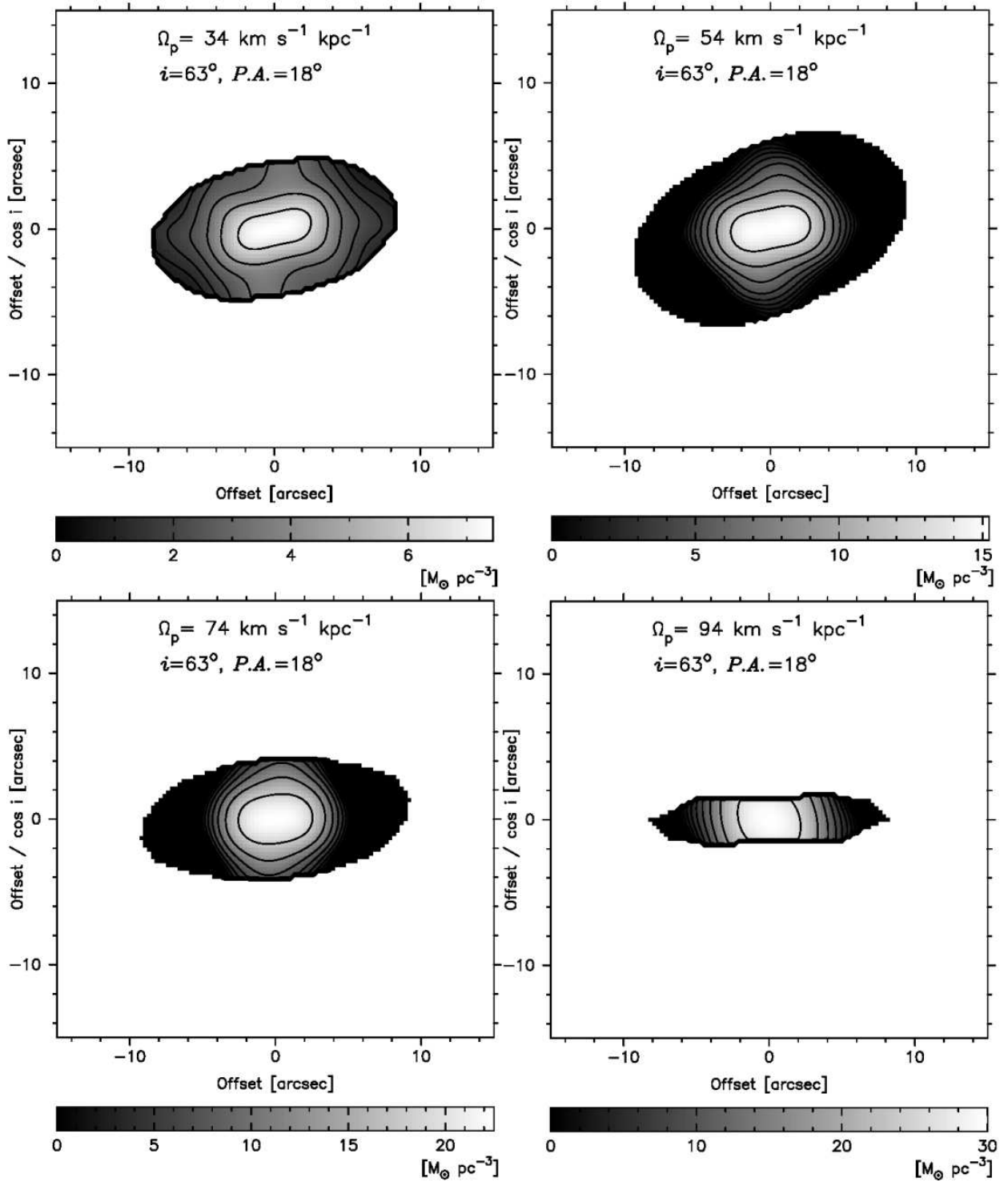}
\caption{Mass distribution for NGC 4569. The contour levels are 26, 18, 13, 9.1, 6.4, 4.5, 3.2, 2.3, and, 1.6 M$_\odot$ pc$^{-3}$. }
\end{figure}


\begin{thebibliography}{}
\bibitem[Binney \& Tremaine(1987)]{bin87} Binney, J.~\& 
Tremaine, S.\ 1987, Princeton, NJ, Princeton University Press, 1987, 747
\bibitem[de Vaucouleurs et al.(1992)]{dev92} de Vaucouleurs, 
G., de Vaucouleurs, A., Corwin, H.~G., Buta, R.~J., Paturel, G., \& Fouque, 
P.\ 1992 (RC3)
\bibitem[Huntley(1978)]{hun78} Huntley, J. M.\ 1978, \apjl, 
225, L101 
\bibitem[Jarrett et al.(2003)]{jar03} Jarrett, T.~H., 
Chester, T., Cutri, R., Schneider, S.~E., \& Huchra, J.~P.\ 2003, \aj, 125, 
525 
\bibitem[Kuno \& Nakai(1997)]{kun97} Kuno, N.~\& Nakai, N.\ 
1997, \pasj, 49, 279 
\bibitem[Schoenmakers et al.(1997)]{sch97} Schoenmakers, R. H. M., Franx, M., de Zeeuw, P. T. 1997, \mnras, 292, 349
\bibitem[Sofue et al.(1999)]{sof99} Sofue, Y., Tutui, Y., 
Honma, M., Tomita, A., Takamiya, T., Koda, J., \& Takeda, Y.\ 1999, \apj, 
523, 136 
\bibitem[Sofue et al.(2003a)]{sof03a} Sofue, Y., Koda, J., 
Nakanishi, H., Onodera, S., Kohno, K., Tomita, A., \& Okumura, S. K.\ 2003a, 
\pasj, 55, 17 
\bibitem[Sofue et al.(2003b)]{sof03b} 
Sofue, Y., Koda, J., Nakanishi, H., \& Onodera, S.\ 2003b, \pasj, 55, 59 
\bibitem[Tully(1988)]{tul88} Tully, R.~B.\ 1988, Cambridge 
and New York, Cambridge University Press, 1988, 221 p.,  
\bibitem[Tremaine \& Weinberg(1984)]{tre84} Tremaine, S.~\& 
Weinberg, M.~D.\ 1984, \apjl, 282, L5 
\bibitem[Wada, Sakamoto, \& Minezaki(1998)]{wad98} Wada, K., 
Sakamoto, K., \& Minezaki, T.\ 1998, \apj, 494, 236 
\bibitem[Wada(1994)]{wad94} Wada, K.\ 1994, \pasj, 46, 165
\end{thebibliography}
\end{document}